\documentclass{elsart}

\usepackage{epsfig}
\usepackage{amssymb,amsmath}

\begin{document}

\begin{frontmatter}

\title{ Impact of observational incompleteness \\
        on the structural properties \\
        of protein interaction networks }

\author[label1,label2]{Mathias Kuhnt}
       \ead{kuhnt@theory.phy.tu-dresden.de}
\author[label3]{Ingmar Glauche}
       \ead{ingmar.glauche@imise.uni-leipzig.de}
\author[label2]{Martin Greiner}
       \ead{martin.greiner@siemens.com}
\address[label1]{Institut f\"ur Theoretische Physik,
                 Technische Universit\"at Dresden,
                 D-01062 Dresden, Germany}
\address[label2]{Corporate Technology, Information \& Communications,
                 Siemens AG, 
                 D-81730 M\"unchen, Germany}
\address[label3]{Institut f\"ur Medizinische Informatik, Statistik
                 und Epidemiologie,
                 Universit\"at Leipzig,
                 H\"artelstr.\ 16/18,
                 D-04107 Leipzig, Germany}

\begin{abstract}
The observed structure of protein interaction networks is corrupted by
many false positive/negative links. This observational incompleteness
is abstracted as random link removal and a specific, experimentally
motivated (spoke) link rearrangement. Their impact on the structural 
properties of gene-duplication-and-mutation network models is studied. 
For the degree distribution a curve collapse is found, showing no 
sensitive dependence on the link removal/rearrangement strengths and 
disallowing a quantitative extraction of model parameters. The spoke
link rearrangement process moves other structural observables, 
like degree correlations, cluster coefficient and motif frequencies,
closer to their counterparts extracted from the yeast data. This 
underlines the importance to take a precise modeling of the 
observational incompleteness into account when network structure
models are to be quantitatively compared to data.
\end{abstract}


\end{frontmatter}

\newpage
\section{Introduction}
\label{sec:intro}

Recent advances in the identification of protein interactions 
\cite{gav02,ho02,ton01} have greatly extended their number in actual 
datasets \cite{sal04,bre03,alf05,zan02}. The accumulated knowledge about
these complex mutual interactions of single proteins is represented in
protein interaction networks where proteins are represented by nodes and 
interactions by links between respective nodes. Investigations on the 
topological structure of the network graphs contribute significantly to 
the understanding of the organizational principles and evolutionary
strategies behind such complex interaction networks.

\begin{figure}
\begin{center}
\epsfig{file=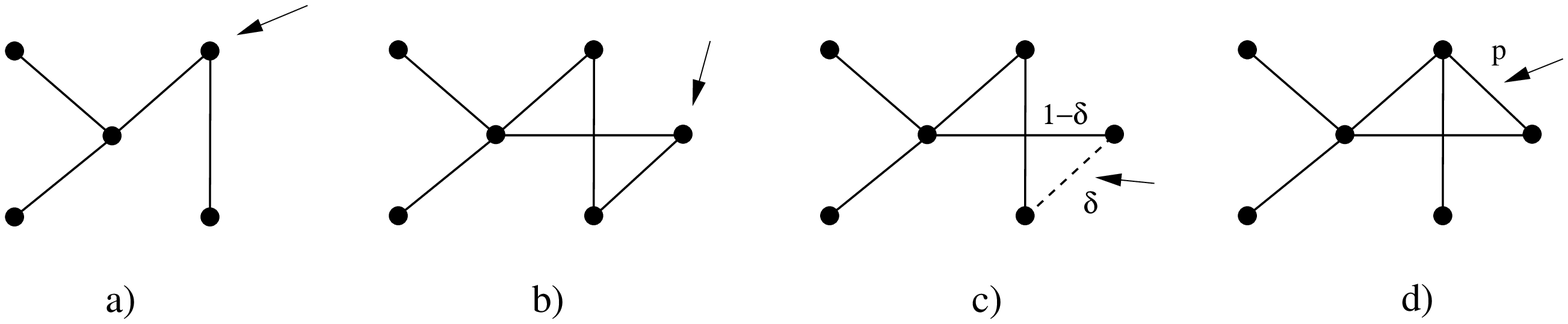,width=0.9\textwidth}
\caption{
The gene-duplication-and-mutation model of Ref.\ \cite{isp05}:
a) random selection of a node,
b) copy of this node with all of its links,
c) deletion of copied links with probability $\delta$,
d) introduction of a new link between original and copied node with
probability $p$.
}
\label{fig:fig1}
\end{center}
\end{figure}

Several models have been proposed for the modeling of the structural 
evolution of protein interaction networks 
\cite{sol02,vaz03,ber04,isp04,isp05}. All of them are based on the idea 
of gene duplication and mutation to be the responsible mechanism for the 
evolution from a small number of proteins up to several thousands known 
today. This mechanism, where links of highly connected nodes are more 
likely to be duplicated, is a biological representation of network growth 
with preferential attachment \cite{eis03}. The model described in 
\cite{isp05} fits best to observed properties of real yeast interaction 
networks, extracted for example from the DIP database \cite{sal04}. 
During one evolutionary step of this model (see Fig.\ \ref{fig:fig1}) a 
randomly selected node is copied with all its links. With probability 
$\delta$ each of the copied links is then subject to removal, and with 
probability $p$ a new (homodimer) link is established between original 
and copied node. If after the probabilistic link removals and addition 
the copied node is left without any link, it is deleted. Fig.\ 
\ref{fig:fig2} illustrates the dependence of the degree distribution on 
the model parameter $\delta$. As for the giant component of real yeast 
data \cite{sal04}, the number of nodes has been set to 
$N_\mathrm{gc} = 4687$. The parameter $p=0.1$ has been estimated 
according to the number of homodimers in real yeast datasets. It has no 
significant influence on the degree distribution. For $\delta = 0.58$ the 
degree distribution matches its data counterpart. Other network 
properties, like degree correlation, cluster coefficient and selected 
motifs, also agree with data to some extend (see Fig.\ \ref{fig:fig4}).

Although the simple gene-duplication-and-mutation mechanism disregards
any selection process and does not take further regulatory mechanisms
into account, it gives us a principle understanding of how evolution
went to work. However, caution should be taken when it comes to a
biological interpretation of the fitted model parameter values. One has
to take into account that the actual data of protein interactions
contains a large number of false links. In Refs.\ 
\cite{bad04,dea02,mer02} different methods are applied to provide an 
estimate about the amount of links which are set in the real yeast
datasets but do not exist (false positives), and those which exist but
are not contained in the dataset (false negatives). Mostly by comparing 
high confidential with high throughput datasets, current estimates are 
that the total number of interactions is up to 30,000 compared to about 
15,000 known today \cite{bad04,mer02} and that within these 15,000 
interactions 50\% of the links are wrongly assigned \cite{dea02,mer02}.

Given this amount of observational incompleteness, two driving questions
emerge: can we compare a model like \cite{isp05} "as is" with data, and
how relevant are fitted model parameters? The aim of this Paper is to
analyze how much and in what directions various forms of observational
incompleteness modify extracted structural properties of protein
interaction networks, like degree distribution, degree correlations,
cluster coefficient and motifs.

In abstracted form, the observational incompleteness leading to the
occurrence of false negative / positive links can be modeled as link
removal / rearrangement applied to an initial ``true'' network. The
simplest variant is completely random removal / rearrangement of links.
Its effect on scale-free networks has already been discussed
\cite{sch05}. Another form of link removal is subnetwork sampling, like
snowball sampling \cite{new03,lee05}, truncated random walk sampling or
traceroute exploration \cite{dal05}.

Throughout this Paper we will make use of the network model of Ref.\
\cite{isp05}. It serves to synthetically generate ``true'' 
protein interaction networks. Sect.\ \ref{sec:two} discusses completely
random deletion of links. Sect.\ \ref{sec:three} introduces a very
specific random link rearrangement, which is directly motivated from the
experimentally applied complex purification methods \cite{gav02,ho02}. A 
conclusion and an outlook is given in Sect.\ \ref{sec:conclusion}.

\section{Random link removal}
\label{sec:two}

Random link removal represents the simplest modeling to introduce false
negative links. A network with $N$ nodes and $L$ links is considered to 
be the initial ``true'' network. One after the other a link is selected
randomly and then removed from the network. The removal strength
$\nu=\Delta{L}/L$ counts the relative number of deleted, i.e.\ false
negative links. The impact of this random link removal on the degree
distribution $p(k)$ of the gene-duplication-and-mutation network of
Ref.\ \cite{isp05} is shown in Fig.\ \ref{fig:fig3}a. With
increasing removal strength the resulting degree distribution deviates
more and more from its initial counterpart.

\begin{figure}
\begin{center}
\epsfig{file=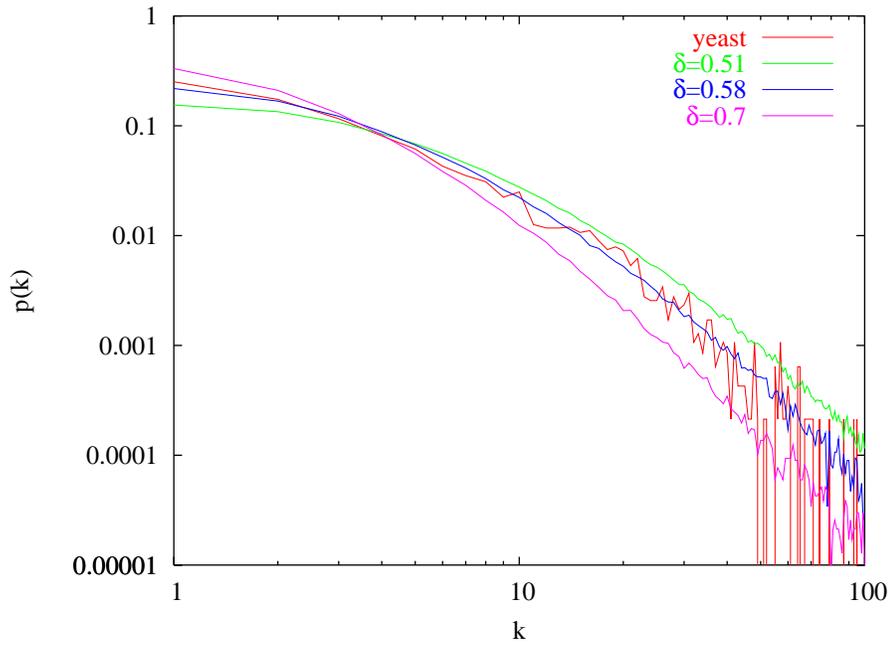,width=0.9\textwidth}
\caption{
Degree distribution resulting from the network model proposed in Ref.\
\cite{isp05} for various parameter values $\delta$. The other
parameters have been set to $N=4687$ and $p=0.1$. The value
$\delta=0.58$ fits best to the yeast protein interaction data taken from
the DIP database \cite{sal04}.
}
\label{fig:fig2}
\end{center}
\end{figure}

\begin{figure}
\begin{center}
\epsfig{file=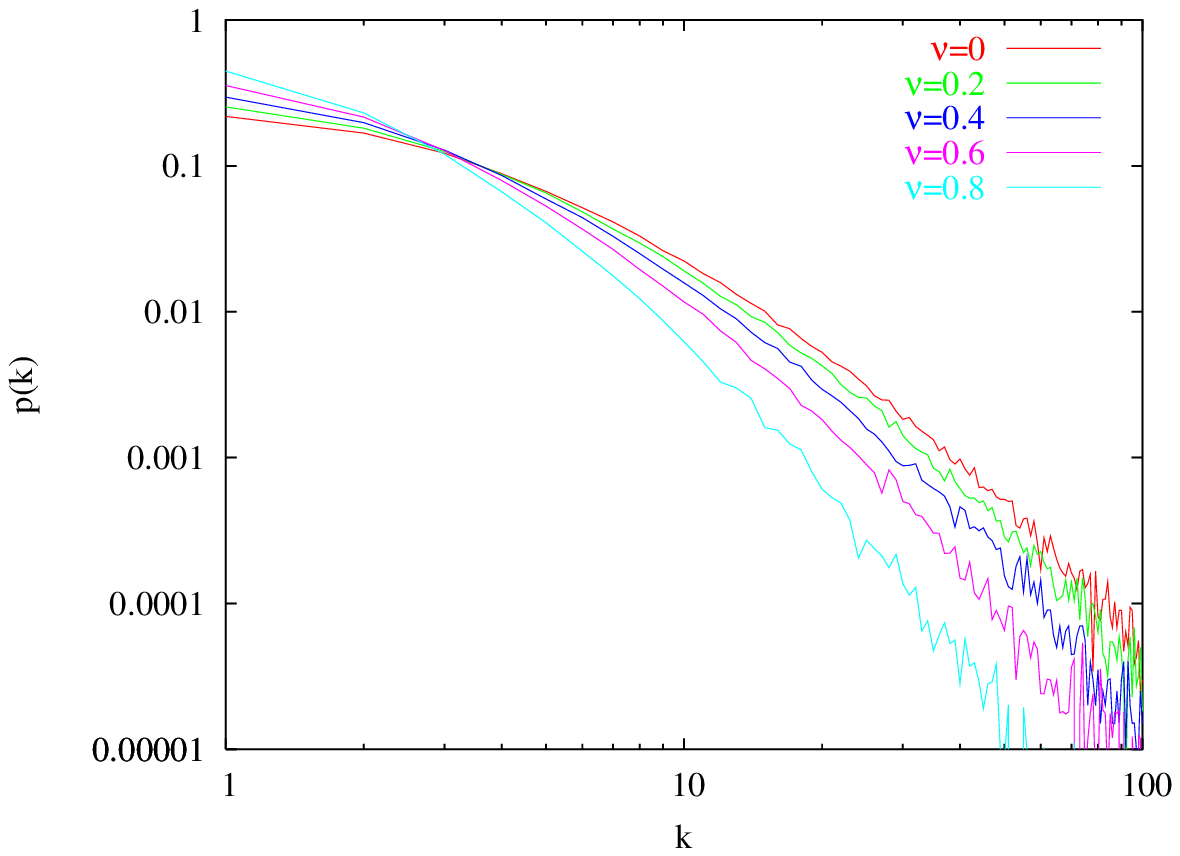,width=0.9\textwidth}
\epsfig{file=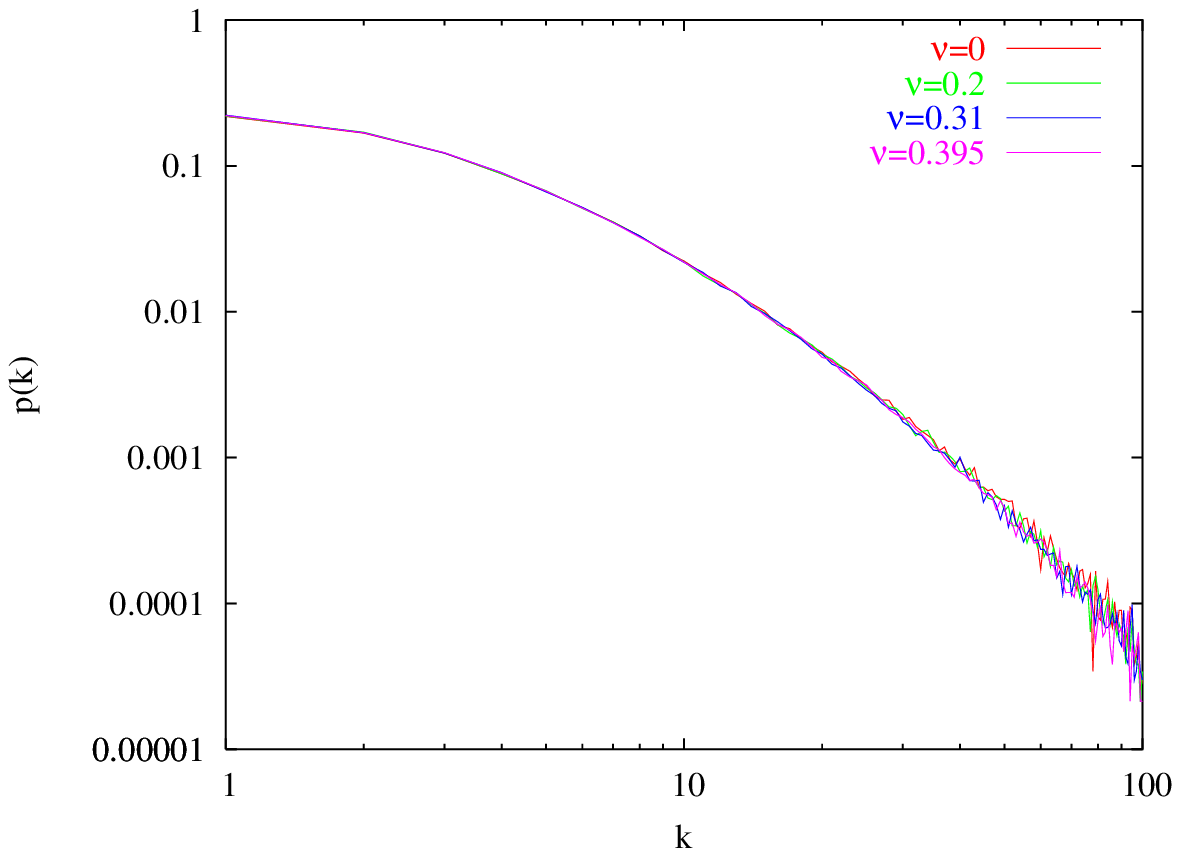,width=0.9\textwidth}
\caption{
Degree distributions for various random link removal strengths $\nu$.
(Top) Parameters of the gene-duplication-mutation model \cite{isp05} 
have been set to $N=4687$, $\delta=0.58$, $p=0.1$, such that 
$\langle k \rangle \approx 6.47$ for $\nu=0$. (Bottom) Model parameters 
have been chosen such that the size of the giant component and the 
average degree become $N_\mathrm{gc} \approx 4687$ and
$\langle k \rangle \approx 6.47$ after link removal; for $\nu = 0.2$,
$0.31$, $0.395$ rescaled parameter values are
$(N,\delta) = (4950,0.55)$, $(5100,0.53)$, $(5250,0.51)$, and $p=0.1$.
The various degree distributions have been sampled from the respective
giant components of $50$ independent network realizations.
}
\label{fig:fig3}
\end{center}
\end{figure}

Admittedly, the degree distributions of Fig.\ \ref{fig:fig3}a with
$\delta=0.58$ and $\nu>0$ resemble those of Fig.\ \ref{fig:fig2} with
$\delta > 0.58$ and $\nu=0$. In fact, the shown degree distributions
with $\nu=0.2$, $0.4$, $0.6$, $0.8$ match those resulting from 
$\delta=0.62$, $0.66$, $0.73$, $0.85$, respectively. By looking at the
degree distribution only, a network with specific $\delta$, but subject 
to random link removal, appears like a network corresponding to a larger
$\delta$. 

Note however, that this comparison is not mature. With $\nu>0$ the size 
$N_\mathrm{gc}$ of the giant network component, from which all degree
distributions of Fig.\ \ref{fig:fig3}a have been sampled, is reduced. 
For $\delta=0.58$ and $\nu=0.2$, $0.4$, $0.6$, $0.8$ it results in 
$N_\mathrm{gc} \approx 4400$, $4000$, $3350$, $2100$, respectively. By 
model construction the initial size $N_\mathrm{gc}(\nu{=}0) = N = 4687$ 
is independent of the parameter $\delta$. Consequently, the number of 
nodes contained in the giant component does not agree between the
link-removed model network and the reparametrized initial model network, 
although their degree distributions match.

For a proper comparison the model network reduced by random link removal
should end up with the same average degree $\langle k \rangle$ and the 
same size $N_\mathrm{gc}$ for the giant component as the reparametrized 
initial network model. For reference, we choose $\langle k\rangle=6.47$
and $N_\mathrm{gc} = 4687$ as observed in the yeast data \cite{sal04}.
This requires the model network to have initially more nodes and links 
before random link removal sets in. Initial numbers of nodes and links 
are not independent of each other and require a careful tuning, so that
after random link removal a precision landing is made at the targeted 
$\langle k \rangle$ and $N_\mathrm{gc}$. For example, for removal 
strengths $\nu = 0.2$, $0.31$, $0.395$ the rescaled parameters are  
$(N,\delta) = (4950,0.55)$, $(5100,0.53)$, $(5250,0.51)$. The remaining 
parameter $p=0.1$ has been kept fixed. Note, that even larger removal 
strengths are not feasible for the chosen network model. It would
require $\delta<0.5$. In this regime the model is not self-averaging any 
longer \cite{isp04}.

Fig.\ \ref{fig:fig3}b illustrates the degree distributions obtained
after random link removal has been applied to the parameter-rescaled 
model realizations. All distributions corresponding to different removal
strengths collapse to one single curve. This curve collapse is somewhat
surprising, because by construction only the size of the resulting giant 
network component and the resulting average degree have been set the 
same. Each curve results from the interplay of two effects: initially, 
i.e.\ before random link removal sets in, a smaller $\delta$ leads to a 
flatter degree distribution (see again Fig.\ \ref{fig:fig2}), which is
then, once random link removal sets in, turned into a steeper 
distribution (see again Fig.\ \ref{fig:fig3}a). 

If the resulting degree distributions had all been Poissonians,
then the curve collapse would have been straightforward to understand. 
The rate equation for random link removal \cite{sch05}
\begin{equation}
\label{eq:two1}
  \frac{d p_k}{d \nu}
    =  \frac{k+1}{(1-\nu)} p_{k+1} 
       - \frac{k}{(1-\nu)} p_{k}
\end{equation}
is solved by $p_k = (\lambda^k/k{!})e^{-\lambda}$ with 
$\lambda = \langle k \rangle = 2L(1-\nu)/N$. A Poissonian degree
distribution remains Poissonian, although with rescaled parameter
$\lambda$. Hence, a Poissonian network constructed with 
$\langle k \rangle = \lambda$ can also be obtained by first constructing
a denser Poissonian network with $\langle k \rangle = \lambda/(1-\nu)$,
which is then subject to random link removal of strength $\nu$.

Also for scale-free distributions $p_k \sim k^{-\gamma}$ the curve 
collapse can be constructed with a rescaling of model parameters. In 
case of a growth process with preferential attachment 
$\pi \sim k{+}\lambda$, the model parameters are the number $m$ of open 
links, with which a new node enters the network, and the attractiveness 
$\lambda$ \cite{alb02}. They determine the scale-free exponent 
$\gamma = 3 + \lambda{/}m$. Now, Ref.\ \cite{sch05} has shown that
during the preferential-detachment-like random link removal, where the 
initial average degree $\langle k \rangle = 2m$ is reduced, the 
scale-free exponent is conserved in the large-$k$ regime. This implies 
that after random link removal with strength $\nu$ the resulting network 
appears like one which has been grown with rescaled model parameters 
$m_\mathrm{rescaled} = (1{-}\nu) m$ and
$\lambda_\mathrm{rescaled} = (m_\mathrm{rescaled}/m) \lambda 
= (1{-}\nu) \lambda$. 

Although the small excursions to Poissonian and scale-free networks have
shed some light on the nature of the curve collapse, its appearance in
connection with gene-duplication-and-mutation networks remains without a 
deeper explanation. Nevertheless, from a pragmatic point of view we can 
say the following: if we consider a gene-duplication-and-mutation 
network as the ``true'' network and introduce false negatives in the 
form of random link removal, then the resulting degree distribution 
appears like one obtained from the same gene-duplication-and-mutation 
process, but with different parameters. It would be inappropriate to 
give a biological interpretation to the magnitude of the extracted
parameters.

\begin{figure}
\begin{center}
\epsfig{file=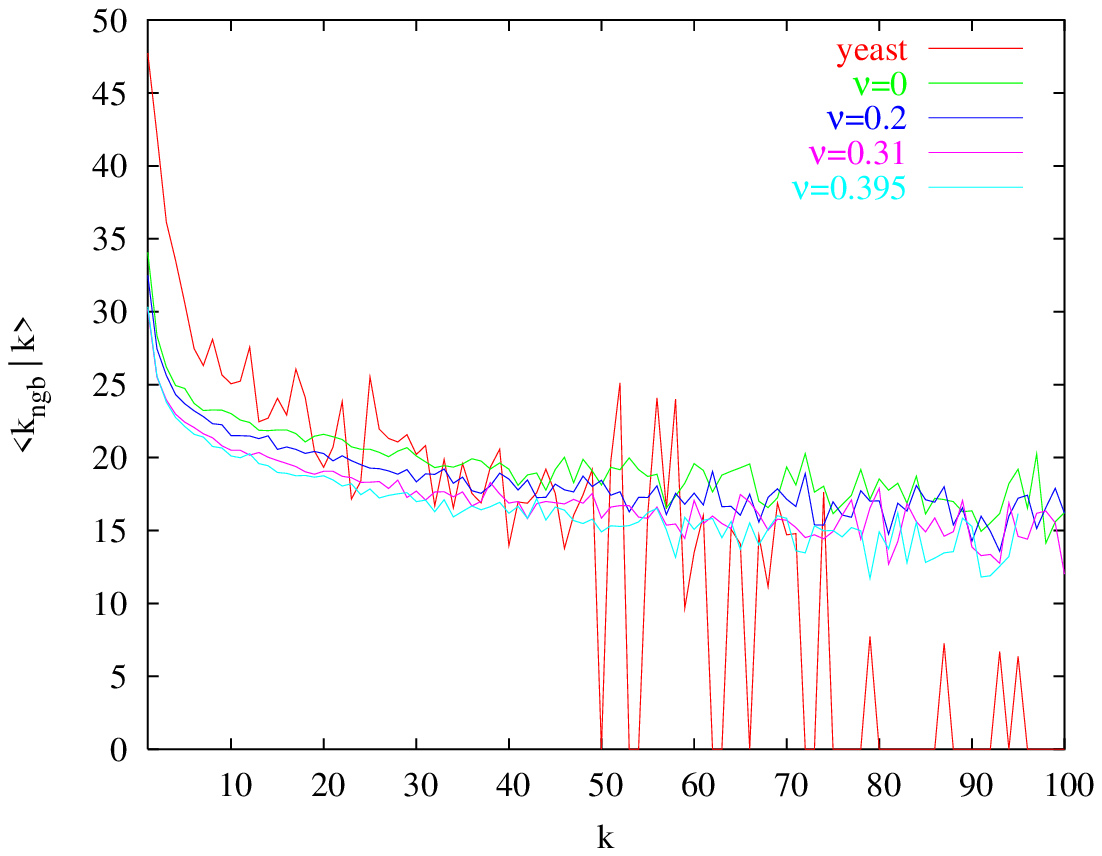,width=0.6\textwidth}
\epsfig{file=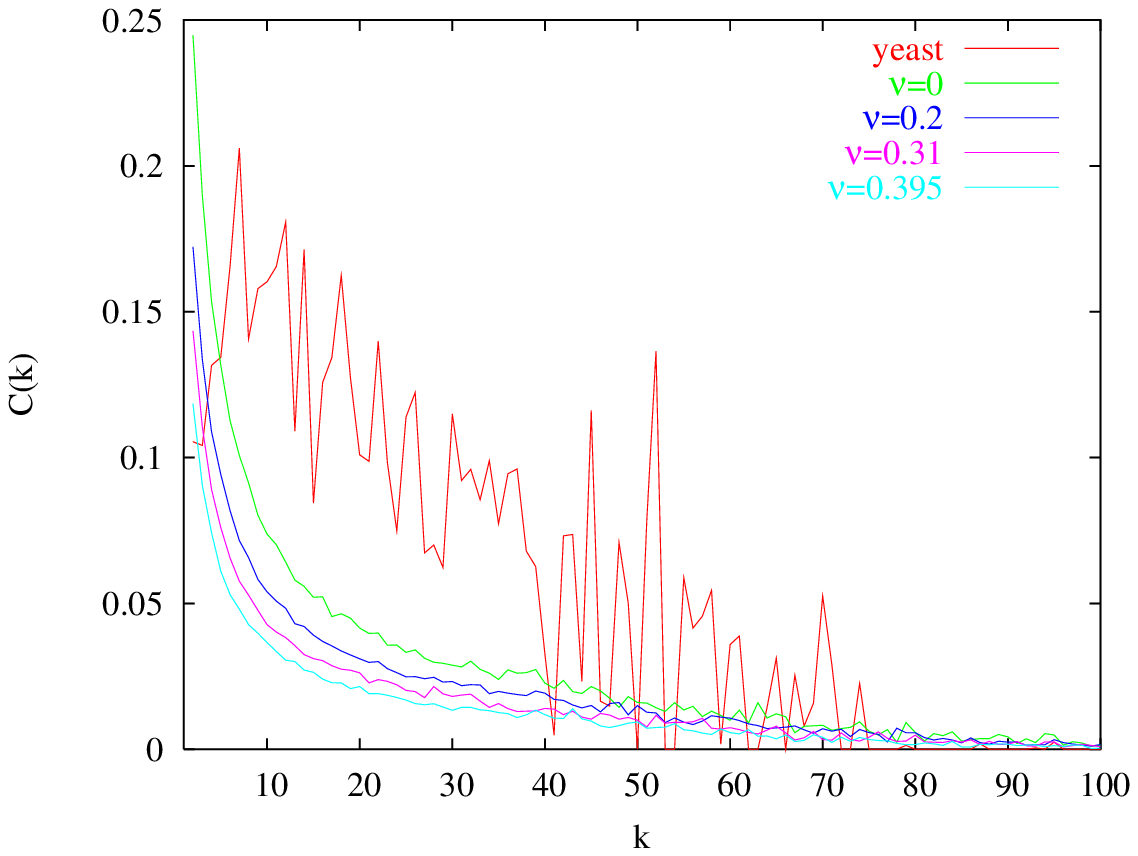,width=0.6\textwidth}
\epsfig{file=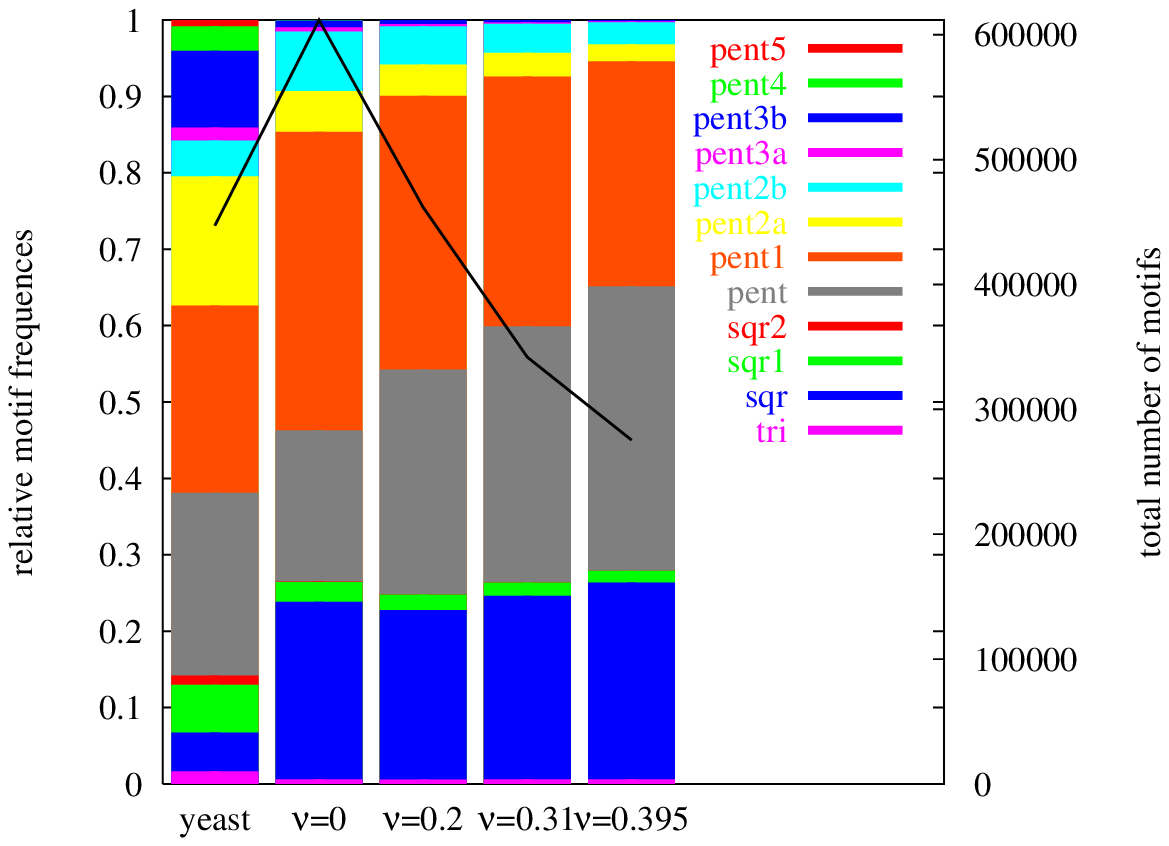,width=0.6\textwidth}
\end{center}
\caption{
(a) Degree correlation, (b) degree-dependent cluster coefficient and (c)
relative frequencies of selected motifs (see Fig.\ \ref{fig:fig5}) for
various random link removal strengths $\nu$. Model parameters are the
same as in Fig.\ \ref{fig:fig3}b. The various distributions have been 
sampled from the respective giant components of $50$ independent network
realizations. For comparison respective distributions obtained from the 
yeast protein interaction database \cite{sal04} are also shown.
\label{fig:fig4}}
\end{figure}

The curve collapse motivates to look at observables beyond degree
distribution. The average degree $\langle k_\mathrm{ngb} | k \rangle$
for neighbors of a node with degree $k$ represents a measure for
degree correlations. Fig.\ \ref{fig:fig4}a illustrates its dependence
on the removal strength. The same procedure with rescaled model 
parameters as for Fig.\ \ref{fig:fig3}b has been applied. With increasing
$\nu$ the degree correlations are reduced to some minor extend. They 
stay close to the $\nu=0$ model correlations. The comparison with the 
correlations observed in the yeast data makes clear that all curves
corresponding to different $\nu$ more or less match with the same 
quality.

A similar finding is obtained for the degree-dependent cluster
coefficient $C(k)$. It represents the fraction of triangles formed by a 
node with degree $k$ and its neighbors out of the maximum possible 
number $k(k-1)/2$. Fig.\ \ref{fig:fig4}b shows its dependence on the
removal strength. $C(k)$ decreases with increasing $\nu$, but remains
within the same order of magnitude as for $\nu=0$. Compared to the
yeast data, all $\nu$ curves are too low and no one of them is really
to be favored over the other ones.

\begin{figure}
\begin{center}
\epsfig{file=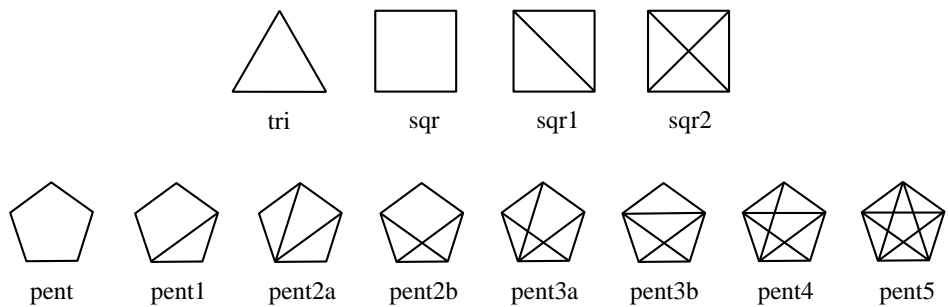,width=0.9\textwidth}
\caption{
Motifs used for the analysis.
\label{fig:fig5}}
\end{center}
\end{figure}

A corresponding conclusion can also be drawn from an analysis based on 
motif structures. A variety of motif systematics has been discussed in 
the literature \cite{mil02,vaz04,prz04,ziv05}. Our selected set is 
depicted in Fig.\ \ref{fig:fig5}. It is restricted to triangles, squares 
and pentagons with different intra-link structure. The loops within 
these motifs represent potential regulatory mechanisms. The total number 
$M = \sum_\mathrm{motifs} M_\mathrm{motif}$ of all selected motifs as 
well as their relative frequencies $M_\mathrm{motif}/M$ have been 
determined in dependence on the random link removal strength. Fig.\ 
\ref{fig:fig4}c reveals that the total number decreases with $\nu$. At 
$\nu\approx0.2$ it matches its yeast data counterpart. Within the model 
the three dominant contributions come from the motifs "sqr", "pent" and 
"pent1". The relative frequency of "sqr" basically remains independent 
of $\nu$, but noticeably overestimates the frequency extracted from the 
yeast data set. With increasing removal strength the relative frequency 
of "pent" increases slightly, whereas that of "pent1" decreases to some 
small extend. Both more or less agree with their yeast data 
counterparts. Except for "pent2b", no agreement is reached for the 
relative frequencies of all "pent" motifs with more than one intra link. 
The significant model underestimations hold for all link removal 
strengths.

\section{Spoke link rearrangement}
\label{sec:three}

So far the modeling of observational incompleteness has only taken
subnetwork sampling in the form of random link removal into account. In 
this way only false negative links have been created. For the modeling 
of false positive links some kind of link rearrangement or link addition 
is needed. We will now discuss a very specific random link 
rearrangement, which is directly motivated from the shortcomings in
the generation and interpretation of protein-interaction data.

\begin{figure}
\begin{center}
\epsfig{file=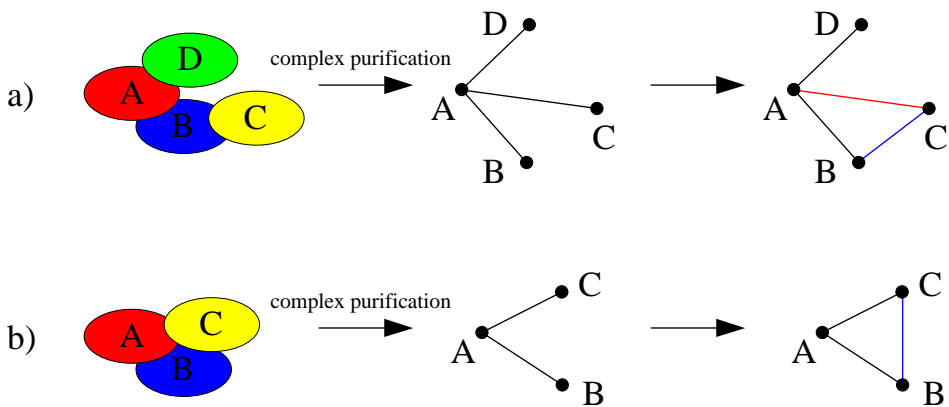,width=0.9\textwidth}
\caption{
Complex purification methods may lead to wrong link assignments.
(a) Bait protein A and prey proteins B, C, D bind for a complex.
Assigned links reflect the bait-prey relationship. However, A does not
directly bind to C (red false positive). It is B, which binds to C (blue
false negative).
(b) For the complex ABC links are only assigned between bait A and preys
B, C. Link B-C is missed, resulting in a (blue) false negative.
}
\label{fig:fig6}
\end{center}
\end{figure}

Using the complex purification methods, namely affinity precipitation 
and affinity chromatography \cite{gav02,ho02}, a protein is tagged and 
placed into the cell lysis. The tagged protein (bait) is then isolated 
and analyzed with its associated proteins (preys). It is not obvious how 
to assign links between the bait and preys found in the protein complex. 
In the commonly used spoke algorithm \cite{bad02} direct links are 
defined between the bait and all its preys. This approach is illustrated 
in Fig.\ \ref{fig:fig6}. It does not take into account the possibility
that the bait is not directly interacting with all preys but via 
intermediate proteins. This results in false positive and negative links 
(Fig.\ \ref{fig:fig6}a). Moreover, possible interactions between the 
prey proteins themselves are also not taken care of, resulting in even 
more false negative links (Fig.\ \ref{fig:fig6}b). Similar effects occur 
with the yeast-two-hybrid \cite{mer02} and the synthetic lethality
methods \cite{ton01}. Although the yeast-two-hybrid method characterizes 
the interaction between two target proteins, no assurance can be given 
that this interaction is not provided by an intermediate protein. With 
the synthetic lethality method an interaction is assumed between two 
functional correlated proteins but even if they are part of the same 
complex, it is not clear if a direct interaction exists.

\begin{figure}
\begin{center}
\epsfig{file=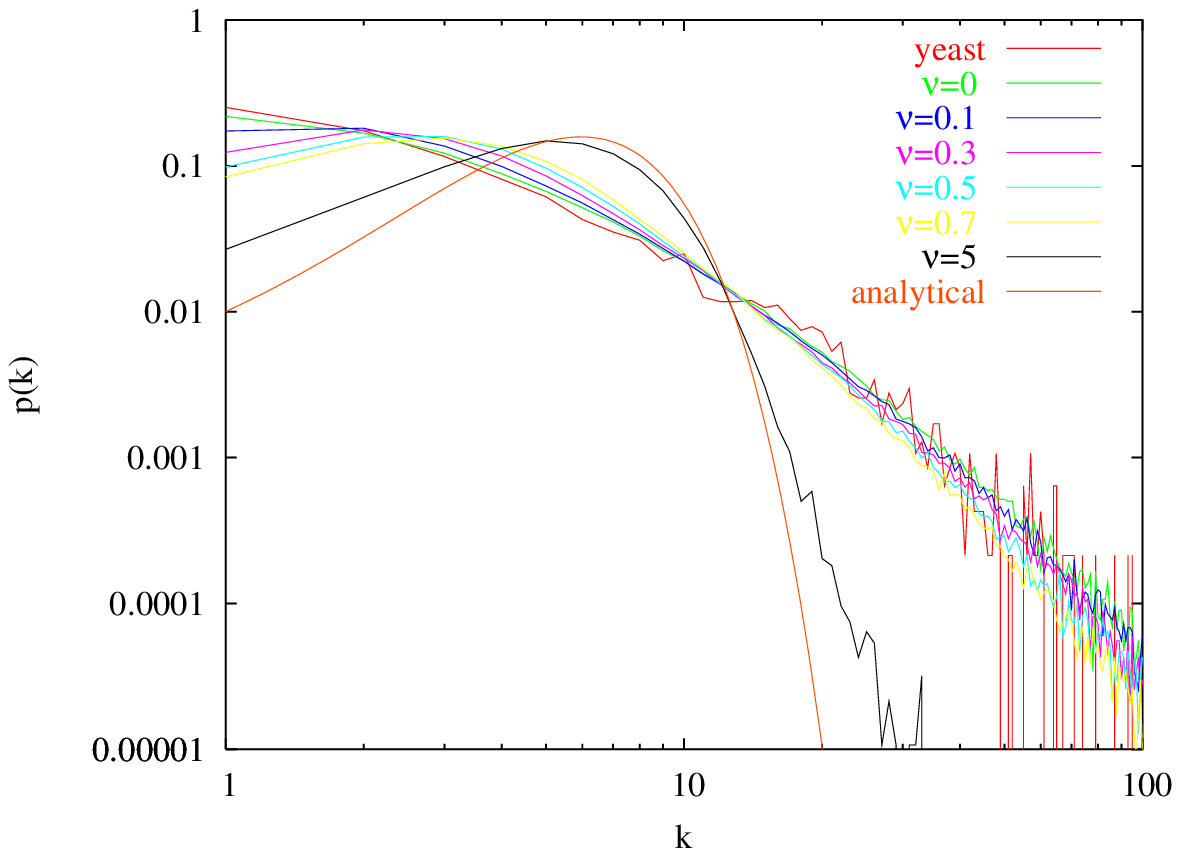,width=0.49\textwidth}
\epsfig{file=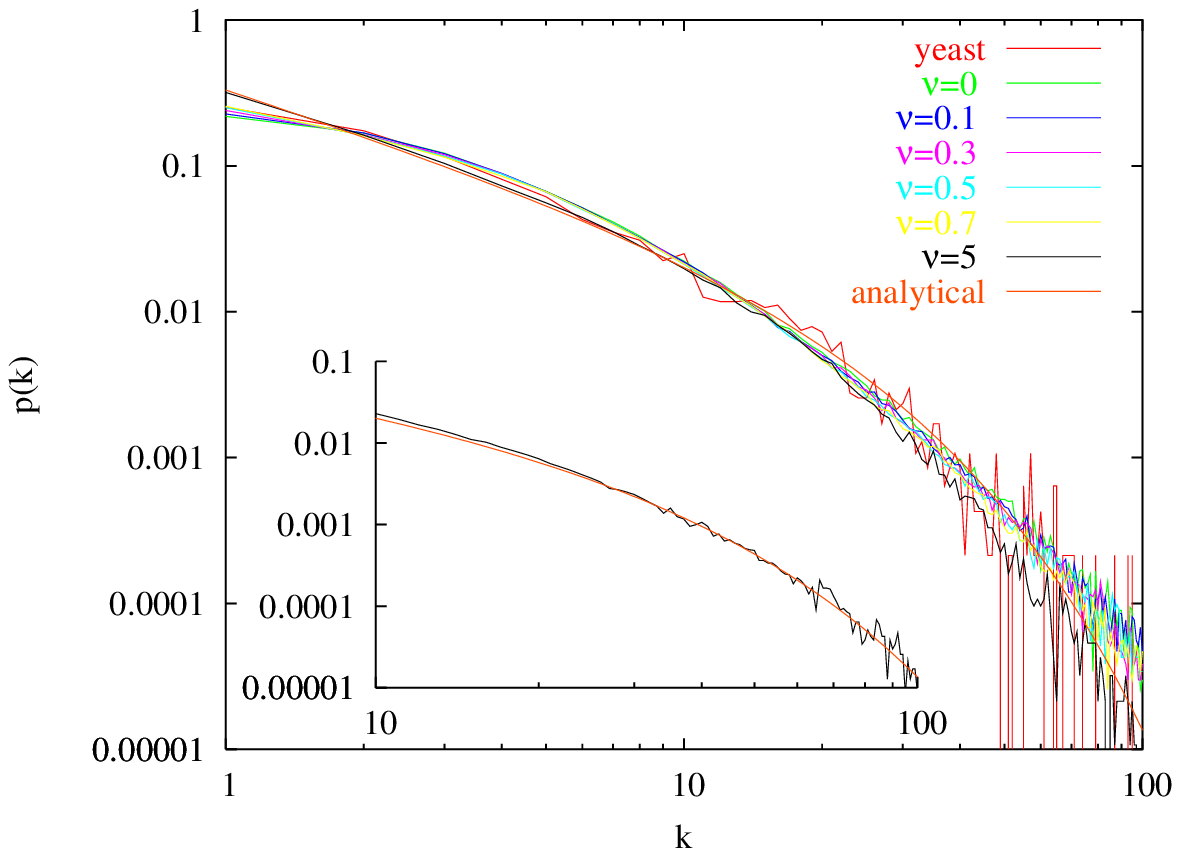,width=0.49\textwidth}
\epsfig{file=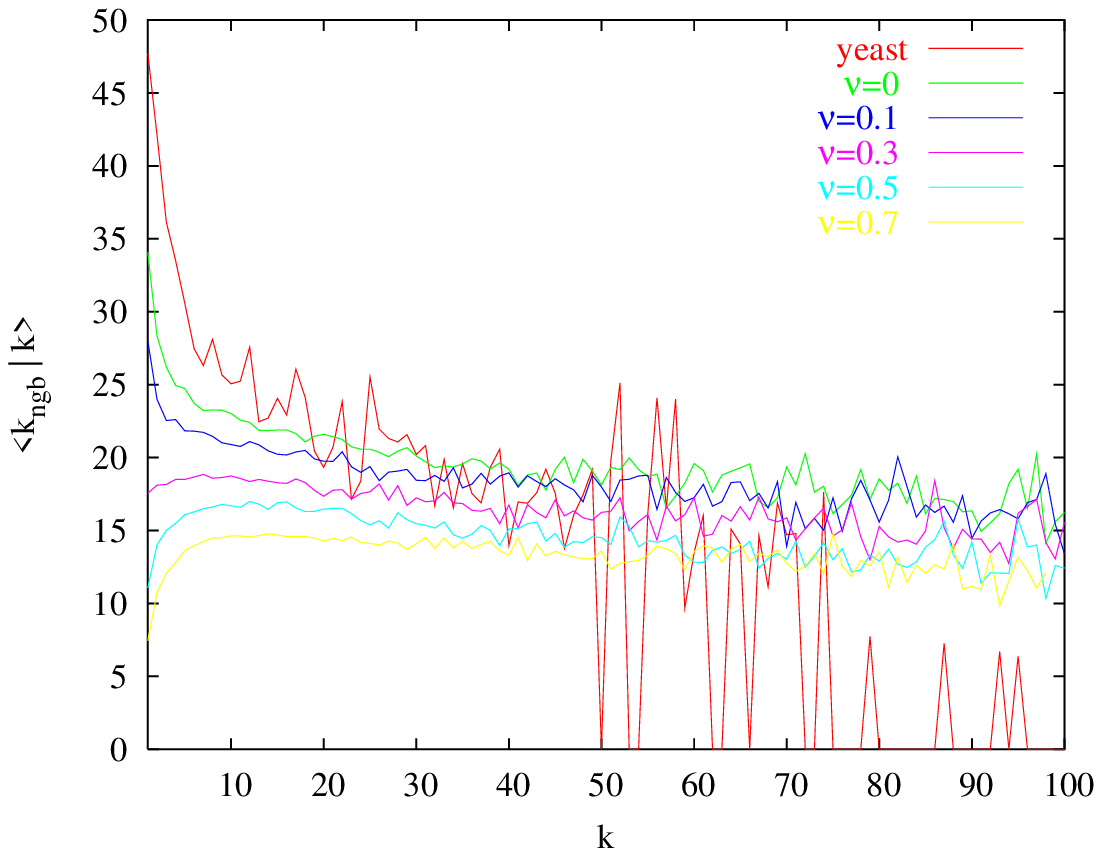,width=0.49\textwidth}
\epsfig{file=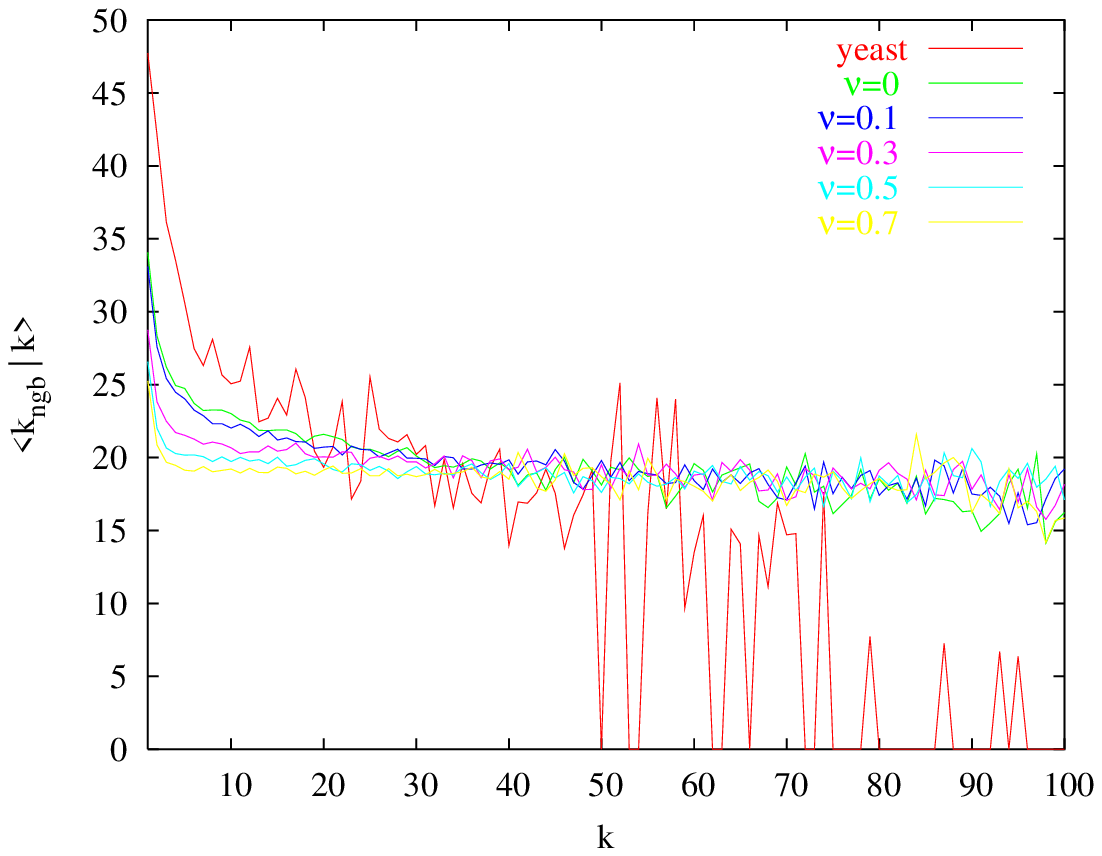,width=0.49\textwidth}
\epsfig{file=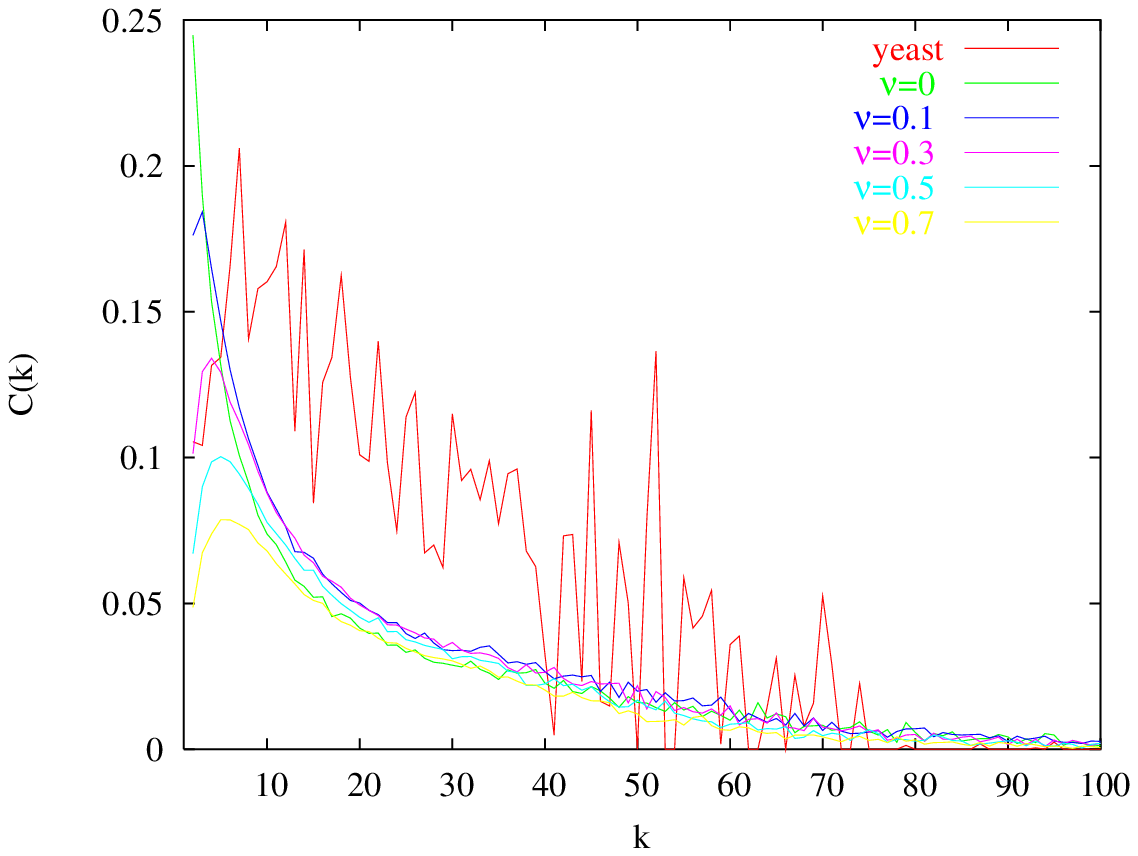,width=0.49\textwidth}
\epsfig{file=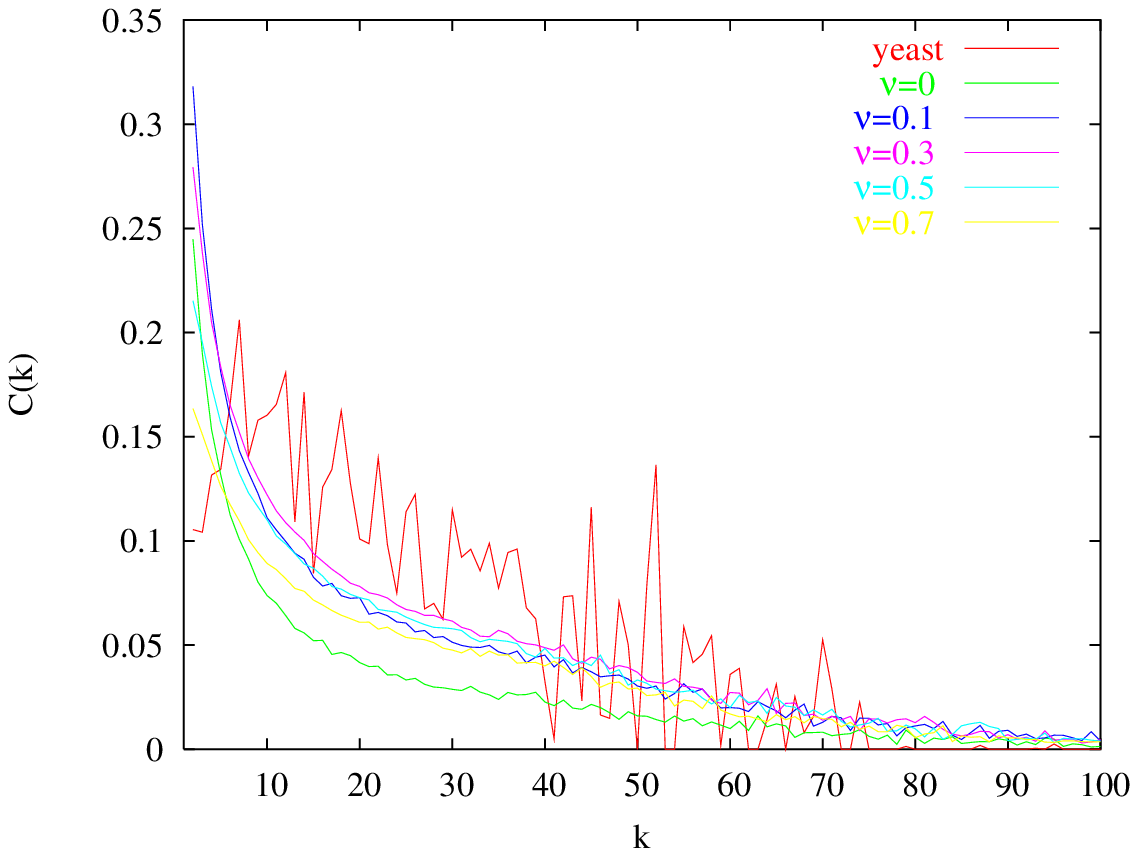,width=0.49\textwidth}
\epsfig{file=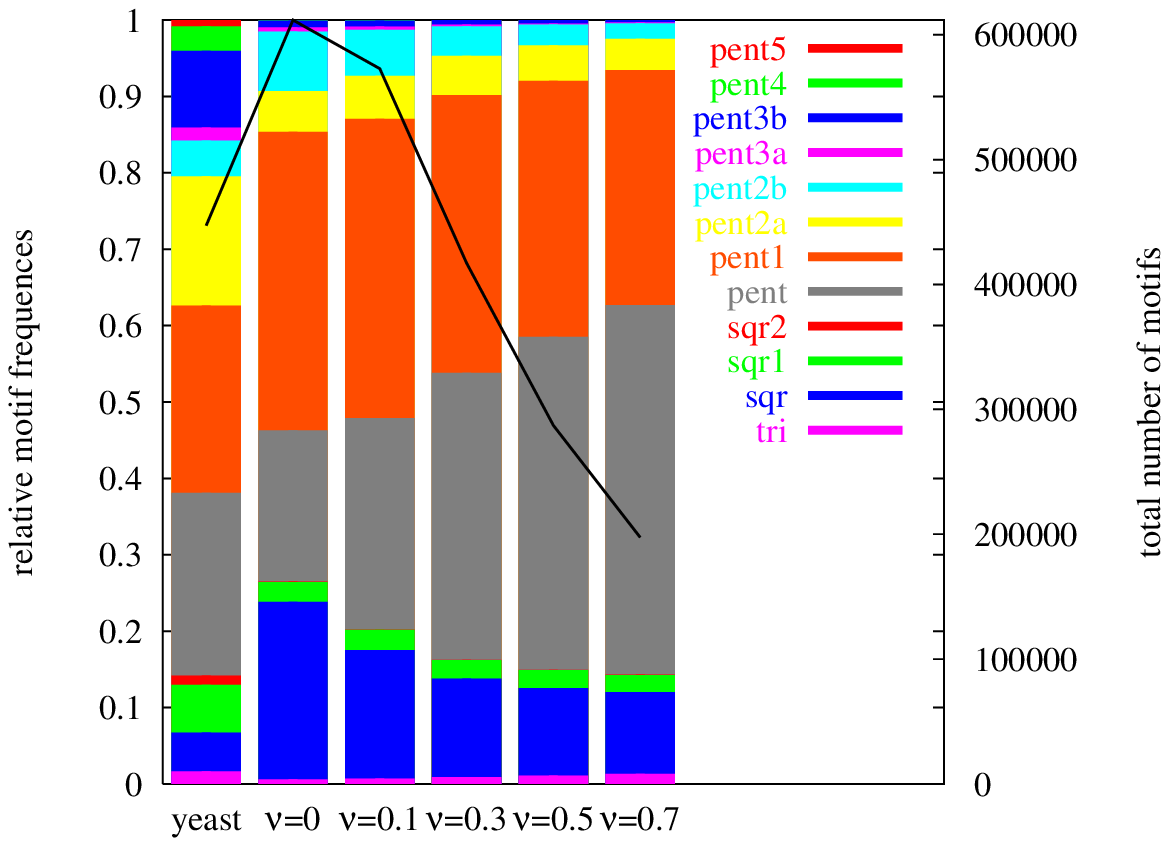,width=0.49\textwidth}
\epsfig{file=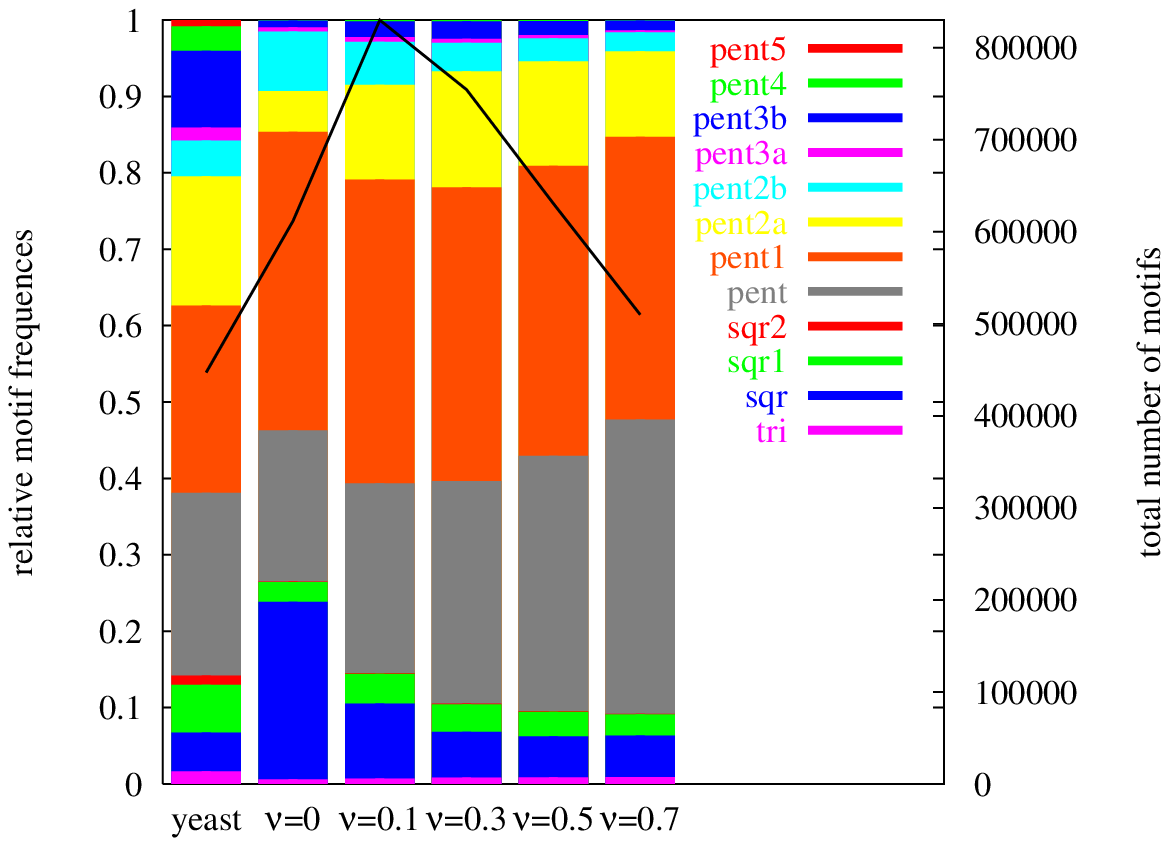,width=0.49\textwidth}
\caption{
(Row 1) Degree distribution, (row 2) degree correlation, (row 3)
degree-dependent cluster coefficient and (row 4) relative motif
frequencies for (left/right column) $\alpha=0$ / $1$ and various spoke
link rearrangement strengths $\nu$. The parameters of the initial
gene-duplication-and-mutation model \cite{isp05} have been set to 
$N=4687$, $\delta=0.58$ and $p=0.1$. The various distributions have been
sampled from $50$ independent network realizations. For comparison, 
respective distributions extracted from the yeast protein interaction 
database \cite{sal04} are also shown. The analytical degree 
distributions (\ref{eq:three4}) and (\ref{eq:three5}), which have been
obtained in the large-$\nu$ limit, are also illustrated in the left and
right part of the first row, respectively.
}
\label{fig:fig7}
\end{center}
\end{figure}

To study the influence of this effect on the network topology we propose
a local random link rearrangement, which hereafter is called spoke link
rearrangement. After selection of an initial (bait) node, one of its
first (prey) neighbors is chosen at random. The latter then continues to
randomly choose one of its first neighbors, excluding of course the
initial node. Two cases then have to be distinguished. If the last node
is a second neighbor of the bait node, a new, but then false-positive
link between these two nodes is introduced and the old link between the
two prey nodes is removed to gain false-negative status; see again Fig.\
\ref{fig:fig6}a. In the other case, the second prey node turns out to be 
a first neighbor of the bait node, upon which only the link between the 
two prey nodes is removed and becomes false-negative; see again Fig.\
\ref{fig:fig6}b.
--
Due to the second case, the spoke link rearrangement is not a pure link
rearrangement. However, the cluster coefficient is small enough to keep 
the link removal part small (see Fig.\ \ref{fig:fig4}b and third row of 
Fig.\ \ref{fig:fig7}).

So far the selection of bait nodes in the spoke link rearrangement has
not been specified. In the yeast data \cite{sal04}, the bait proteins
are of course known and make up approximately a quarter of all listed
protein nodes. In general they have a larger degree than the overall
average. Their degree distribution $p_\mathrm{bait}(k)$ is different
from the observed overall degree distribution $p(k)$, but can be mapped 
onto the latter via
\begin{equation}
\label{eq:three1}
  p_\mathrm{bait}(k)
    \sim  k^\alpha p_{k}
\end{equation}
with $\alpha \approx 0.3$. This indicates that a bait node $i$ with
degree $k_i$ might be picked from the model network with the 
preferential bias
\begin{equation}
\label{eq:three2}
  \Pi_\mathrm{bait}(k_i)
    =  \frac{k_i^\alpha}{\sum_{j=1}^N k_j^\alpha}
       \; . 
\end{equation}
Since the observed degree distribution $p_k$ entering (\ref{eq:three1})
is most likely not equal to the unknown true one, we will discuss 
probabilistic bait selection with $\alpha=0$ and $1$ in the following.

The combination of the biased bait selection (\ref{eq:three2}) and the
spoke link rearrangement process are applied to the network structure
obtained with the gene-duplication-and-mutation model of Ref.\
\cite{isp05}. Again, model parameters are taken to match the yeast
data, i.e.\ $N=4687$, $\delta=0.58$ and $p=0.1$. The rearrangement
strength $\nu=\Delta{L}/L$ counts the relative number of bait selections
implying link rearrangement or removal.

Since link removal is included in the spoke link rearrangement, the
average degree decreases with increasing $\nu$ from its initial value
$\langle k \rangle = 6.47$. For the already very large rearrangement
strength $\nu=0.8$ we arrive at the slightly reduced values
$\langle k \rangle = 6.40$ and $5.97$ for $\alpha=0$ and $1$, 
respectively. Note also, that the giant component of the network does 
not change with $\nu$ and remains at its initial value 
$N_\mathrm{gc}=N$. Both subprocesses of the spoke link rearrangement 
always keep the three involved nodes connected to the overall network. 
This observation together with the decreasing average degree would 
imply, that for a fair comparison between the yeast data and the 
spoke-link-rearranged model a reparametrization of the model parameter 
$\delta$ is required to start with an initially denser network. Since 
the reduction of the average degree remains rather small for modest,  
data-relevant rearrangement strengths, we abandon to do so.

The first row of Fig.\ \ref{fig:fig7} shows the dependence of the degree
distribution on the rearrangement strength $\nu$. For $\alpha=0$ the 
probability to find low- and high-degree nodes decreases with increasing 
$\nu$. Already for small $0 \leq\nu\leq 0.5$ the deviations to the 
initial degree distribution are noticeable. For very large $\nu$ the 
degree distribution appears to converge towards a Poissonian. For 
$\alpha=1$ the outcome is different. For small $\nu$ the deviations to 
the initial degree distribution are barely noticeable, leading to a 
curve collapse in good approximation. Even for large $\nu=5$ the 
resulting $p_k$ is still close by. 

Analytical insight into these findings can be obtained from the 
following rate equation:
\begin{eqnarray}
\label{eq:three3}
  \frac{2}{\langle k \rangle} \frac{d p_k(\nu)}{d \nu}  
    &=&  (1-\delta_{k1}) 
         \frac{(k-1)^\alpha}{\langle k^\alpha \rangle} p_{k-1}(\nu)
         - (1-\delta_{k0})
         \frac{k^\alpha}{\langle k^\alpha \rangle} p_k(\nu)
         \nonumber \\
    & &  + (1-\delta_{k0})
         \frac{k+1}{\langle k \rangle} p_{k+1}(\nu)
         - (1-\delta_{k1})
         \frac{k}{\langle k \rangle} p_{k}(\nu)
         \; .
\end{eqnarray}
The first two terms on the right-hand side represent the gain and loss
term of the selected bait, which increases its degree by one. The third
and fourth term describe the first neighbor of the bait, which looses a
link. Consult again Fig.\ \ref{fig:fig6}a. The spoke link removal (see 
Fig.\ \ref{fig:fig6}b) has been neglected in (\ref{eq:three3}). 
Furthermore, also degree correlations have been discarded.

For the case $\alpha=0$ the stationary solution of (\ref{eq:three3}) is 

found to be the modified Poissonian
\begin{equation}
\label{eq:three4}
  p_k  
    =  \begin{cases}
         \frac{\langle k \rangle^k}{k!} 
         \frac{1}{e^{\langle k \rangle}-1} 
         & (k\geq1) \\
         0 & (k=0)
       \end{cases}
       \; ,
\end{equation}
where $p_0=0$ is required from $N_\mathrm{gc} = N$ for all $\nu$. This 
confirms the simulational finding of Fig.\ \ref{fig:fig7}a for very 
large $\nu$.
--
For $\alpha=1$, $p_k \sim k^{-1}$ represents the non-normalizable 
stationary solution of the rate equation. However, due to the finiteness 
of the network a cutoff $k_c$ may be introduced, leading to
\begin{equation}
\label{eq:three5}
  p_k  =  \begin{cases}
            a k^{-1} e^{-\frac{k}{k_c}}
            & (k\geq1) \\
            0 & (k=0) \; .
          \end{cases}
\end{equation}
With parameters $a=0.35$ and $k_c=18$ this solution is also illustrated 
in Fig.\ \ref{fig:fig7}b and its inset. It agrees nicely with 
$p_k(\nu=5)$ obtained from the simulations.

The degree correlation in the form $\langle k_\mathrm{ngb}|k \rangle$ is
illustrated in the second row of Fig.\ \ref{fig:fig7}. In case of
$\alpha=0$ and especially for very small degrees, the average neighbor
degree rapidly decreases with increasing rearrangement strength. Its 
initial disassortative character is turned into a randomized one, which 
is reflected in the $k$-independence. The average neighbor degree 
associated with $\alpha=1$ does show a similar, but weaker dependence on 
the rearrangement strength.

The third row of Fig.\ \ref{fig:fig7} focuses on the degree-dependent
cluster coefficient. Not much happens for $\alpha=0$ at modest 
rearrangement strengths $\nu \leq 0.7$, except for very low degrees, 
where the cluster coefficient decreases as $k \to 0$. This behavior is 
also observed in the yeast data. In case of the biased bait picking with 
$\alpha=1$, the cluster coefficient increases for all $k$ as the 
rearrangement strength increases from $\nu=0$ to $\nu\approx 0.3$, only 
then to decrease again for even larger $\nu$. At the turning point 
$\nu\approx 0.3$, the found degree-dependent cluster coefficient almost 
matches its counterpart from the yeast data.
--
The overall cluster coefficient declines with increasing 
$\nu=0.1,\,0.3,\,0.5,\,0.7,\,5$ as 
$\langle C \rangle=0.14,\,0.11,\,0.08,\,0.06,\,0.002$ for $\alpha=0$ and 
as $\langle C \rangle=0.20,\,0.19,\,0.16,\,0.12,\,0.008$ for $\alpha=1$. 
For very large rearrangement strengths it becomes very small. 

The motifs of Fig.\ \ref{fig:fig5} are exemplified in the last row of 
Fig.\ \ref{fig:fig7}. For $\alpha=0$ their total number is a strictly
decreasing function with $\nu$. The case $\alpha=1$ shows a different 
behavior. For $\nu=0$ to about $\nu\approx 0.2$ it is first an 
increasing function and then becomes a decreasing function beyond this 
point. Compared to the yeast data, the order of magnitude is right for 
both $\alpha$-values. The relative frequency of the motif 'sqr' 
decreases with $\nu$. Characteristic trends are also observed for other 
motifs, but it is difficult to provide a solid explanation for it. Just 
by looking at the two subfigures, we have the impression that for the 
combination $\alpha=1$, $\nu\approx 0.3$ the distribution of relative 
motif frequencies comes closest to the yeast distribution.

We have arrived at a remarkable result: in comparison to the respective
yeast-data counterparts, the degree distribution resulting from the
parameter combination $\alpha=1$ and $\nu\approx 0.3$ of the spoke link
rearrangement perfectly matches, the found degree correlation is at
least close, the degree-dependent cluster coefficient matches close to
perfect and also the distribution of relative motif frequencies comes
very close. Compared to the initial model at $\nu=0$, the agreement with
data has improved.

Without showing, we have also looked at a combined application of spoke
link rearrangement and random link removal. Results turn out to be a
mere superposition of those obtained independently in this and the
previous section.

\section{Conclusion}
\label{sec:conclusion}

Observed protein interaction networks are known to be corrupted by a
large amount of false negative and false positive links. This
observational incompleteness impacts the analysis of network structure.
We have assumed the emergence of false negative links to be random and
have modeled it with a random subnetwork sampling like random link
removal. Most of the false positive links arise due to an operationally
defined link assignment during measurements. The latter has been
abstracted as a specific random (spoke) link rearrangement process. The
modeling of both forms of observational incompleteness reveals that the
resulting degree distributions either do not depend at all or only weakly 
on the applied link removal / rearrangement strengths. Based on this 
curve collapse alone, no judgment can be made on the qualities of the 
underlying gene-duplication-and-mutation network models and no biological 
interpretation should be given to respective model parameters, like the 
mutation rate $\delta$.

For observables beyond degree distribution, like degree correlation, 
cluster coefficient and motif frequencies, a dependence on the applied
link removal / rearrangement strength is found. Whereas for random link
removal this dependence remains small, spoke link rearrangement appears 
to move these observables closer to their counterparts extracted from 
the DIP database. This shows the importance to include observational 
incompleteness into the comparison between network models and data. It 
should also be included into any systematic identification of 
statistically significant network measures \cite{ziv05} and will gain 
more interest, the more rigorous the analysis of relevant data becomes.

\newpage

\newpage


\end{document}